**The role of graphitic filaments in resistive switching behaviour of amorphous silicon carbide thin films**

P.Chaitanya Akshara[1], Nilanjan Basu[2], Jayeeta Lahiri[2], G.Rajaram[1,2] and M.Ghanashyam Krishna[1,2,*]

[1]Centre for Advanced Studies in Electronics Science and Technology, [2]School of Physics, University of Hyderabad, Hyderabad-500046, Telangana, India
*Corresponding Author: email: mgksp@uohyd.ernet.in

**Abstract**

Resistive switching in amorphous silicon carbide (a-SiC) films deposited by a single composite target magnetron sputtering process is reported. Switching performance as a function of thickness of the films (50, 100 and 300 nm) as well as different top metal electrodes (Cu, Pt and Ag) with the bottom electrode fixed as Au, is investigated. The switching parameters (Forming Voltage, Set and Reset voltages and corresponding currents) are found to be dependent on thickness of SiC films and it is observed that 100 nm is the optimal thickness for best endurance. The interface between metal electrode and a-SiC films plays a more significant role in achieving switching performance. Resistance Off/On ratios of $10^8$, retention times >$10^4$ s and endurance of 50 cycles are achieved in the best devices. Cross-sectional scanning electron microscopy provides evidence that the mechanism of switching involves the formation of carbonaceous filaments and Raman spectroscopy indicates that these filaments are nanocrystalline graphite in nature. The current work clearly establishes that there is dissociation of SiC during the switching cycles leading to formation of nanocrystalline graphitic filaments. These contribute to switching, in addition to the metallic filaments, in the a-SiC based resistive memory device.

**INTRODUCTION**

Resistive random access memories (RRAMs), based on resistive switching in dielectric and semiconductor materials, have been the subject of intense research because of the several advantages they offer over other non-volatile RAMs. While the majority of the work is focused on oxide based RRAMs, [1-6] there is also a large body of literature non-oxide based RRAMs,[7-14] clearly indicating the versatility of the technology. SiC occurs in different polytypes which are particularly attractive for high temperature and high power electronics applications. In the last few years, SiC has also been studied as an active switching material for this technology.[9-12] The ability to use SiC as the active material for RRAM technology is expected to provide several advantages related to compatibility with existing CMOS processing technologies and high temperature applications. Recent studies have demonstrated that SiC based RRAMs exhibitultrahigh resistance ratios in the range of $10^8$-$10^9$ and it is also possible to use them as a multibit storage device. Most importantly, in contrast to the oxide based RRAMs, there is no need to control the oxygen content which leads to oxygen diffusion into the metal electrodes. This is another factor that makes SiC based RRAM important for future technologies.

It has been shown by earlier workers[10,12] that amorphous SiC, rather than crystalline SiC, is required for resistive switching. The device fabricated on Thermally oxidized Si substrate consisted of an amorphous layer of SiC sandwiched between the Au bottom electrode and Cu top electrode.[10] The authors have suggested that the formation of Cu filaments in the ON state could be the origin of conduction in these devices. However, dependence of switching on different top electrodes and the possibility of existence of other mechanisms of reversible switching in SiC has not been investigated. The aim of the current work is to address these questions. The SiC films were prepared by a single composite target RF magnetron sputtering process reported recently, by some of the current authors.[15] The films were crystalline on Si and amorphous on glass as described in the previous work. In contrast to earlier work, RRAMs on glass substrates are demonstrated. In order to compare the performance of devices fabricated in the present study with literature, only the amorphous films were considered. In this work the bottom electrode in all the fabricated devices is Au while the top electrode is Cu, Pt or Ag. The top electrode is changed to study the process of electroforming in these devices. A detailed understanding of the mechanism of switching in these devices is achieved using Raman and x-ray photoelectron spectroscopy. It is demonstrated that formation of graphitic filaments plays a very important role in determining switching performance.

**EXPERIMENTAL DETAILS:**

Three device structures, Au/SiC/Pt, Au/SiC/Cu and Au/SiC/Ag were fabricated on glass substrates. The SiC switching layer was deposited by single composite-target RF magnetron sputtering as described in detail in an earlier study[15]. Briefly, the target consisted of Graphite pieces placed on a Silicon base and the carbon content in the film was controlled by varying the graphite coverage area. The conditions optimized earlier to achieve stoichiometric SiC are used in the present study to fabricate the devices. The thickness of the SiC films was varied from 50 nm to300 nm. The Au bottom electrode and Cu top electrode were deposited by thermal evaporation while the Pt top electrode was deposited by RF sputtering. The Cu and Pt electrodes were of 40μmx40μm size while Ag dots of 500μm diameter were fabricated by using shadow mask.

Au/SiC/Pt and Au/SiC/Cu structures were patterned using photolithography which employed three masks for deposition and lift-off of each layer. The current–voltage (I–V) characteristics, endurance and retention time of the structures were measured at room temperature in air using

Agilent B1500A semiconductor Device Analyzer by grounding the bottom electrode Gold (Au) and applying voltage sweep to the top electrode.

Thickness of the films was measured in a AMBIOS XP 200 profilometer. The amorphous nature of the SiC films was established through X-Ray Diffraction(XRD)measurements obtained at a grazing incidence angle of 0.5º using Bruker D8 Discover X-ray Diffractometer with Cu-K$_\alpha$ radiation (λ=1.54056Å).The cross-sectional microstructure of the devices was imaged in a Field emission-Scanning Electron Microscope (FE-SEM Carl ZEISS, Ultra 55).The X-Ray Photoelectron Spectroscopy (XPS) spectrum was recorded using Kratos Axis Ultra DLD system equipped with monochromatic Al Kα X-ray source at X-ray energy 1486.6 eV. The energy of photoelectrons were measured with a resolution of 0.1 eV. The photoelectron energy spectra were analyzed using CASA XPS software.The Raman spectra of the samples were recorded at an incident wavelength of 532 nm using Witec alpha 300 spectrometer.The D and G peaks were fitted using Lorentzian Function after background subtraction.

The characteristics of the following devices are described in the paper (in each case three devices were tested)

- D1:Au/SiC(50nm)/Ag
- D2:Au/SiC(100nm)/Ag
- D3:Au/SiC(300nm)/Ag
- D4:Au/SiC(50nm)/Cu
- D5:Au/SiC(100nm)/Cu,
- D6:Au/SiC(50nm)/Pt
- D7:Au/SiC(100nm)/Pt

XRD patterns of D5 and D7 shown in Fig.1(a) and 1(b) consist of peaks corresponding to the metal electrodes alone. This indicates the SiC films are amorphous, consistent with the earlier study[15].

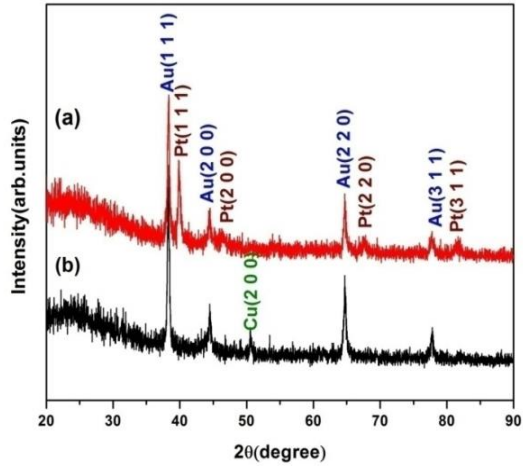

**Figure 1.** X-ray diffraction patterns of devicesbefore testing for resistive switching (a) Au/SiC(100nm)/Pt, (b) Au/SiC(100nm)/Cu

**RESULTS AND DISCUSSION:**

The forming process is the critical first step in all resistive switching devices, since the magnitude of the forming voltage determines the (performance) I-V characteristics of the device. In oxide thin film based RRAMs, it has been shown that there is a weak dependence of the forming voltage on thickness of the active layer but stronger dependence on the top electrode[16]. The current-voltage (I-V) curves for the SiC films in devices D2,D5,D7 are shown in Fig. 2(a)-(c) for different top electrodes of thickness 100nm. I-V curves for the devices D1,D3,D4,D6 are shown in Fig. S1 (a)-(d). Only the first three cycles are shown in each case even though the experiments were carried out over a large number of cycles.The maximum number of cycles before device failure occurs(endurance) in each case will be discussed later.

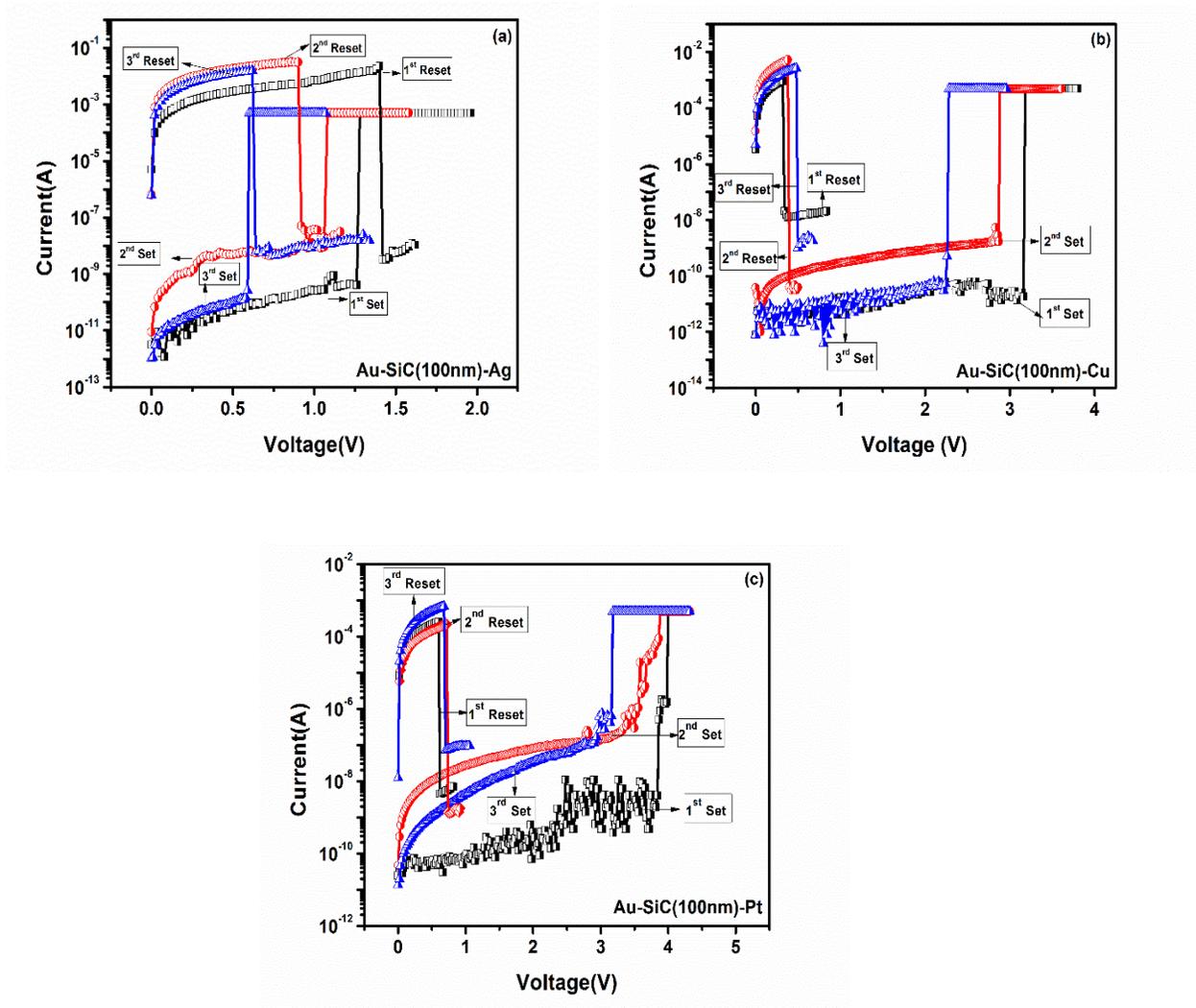

**Figure 2.** Switching characteristics of devices in the first 3 cycles (a) Au/SiC(100nm)/Ag, (b) Au/SiC(100nm)/Cu, (c) Au/SiC(100nm)/Pt

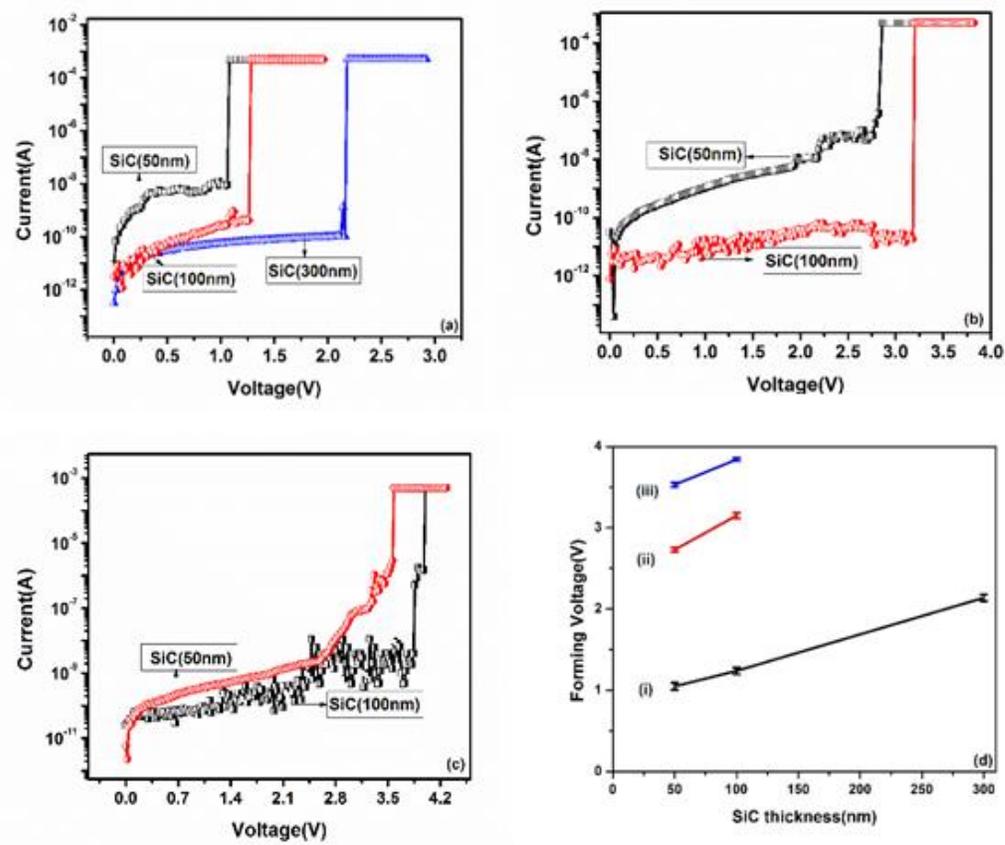

**Figure 3.** Variation in forming voltage as a function of SiC film thickness for different top electrodes (a) Silver (b) Copper (c) Platinum(d)Comparison of Variation in Forming Voltage with respect to change in thickness of SiC and change in Top electrode (i) Au/SiC/Ag, (ii) Au/SiC/Cu, (iii) Au/SiC/Pt

The forming voltages are 1.05V,1.24V,2.14V for D1,D2,D3 which have the same top electrode Ag,with SiC thicknesses 50nm,100nm,300nm respectively. The corresponding values are 2.73V,3.15V for devices having Cu as top electrodes with the SiC thicknesses of 50nm,100nm and 3.53V,3.84V are the forming voltages for devices having Pt as top electrodes with the same SiC thicknesses.The forming voltages are slightly higher than in previous reports which maybe due to the higher thickness of the SiC films in the present case[10]. The thickness of SiC here is 100 nm as against 40 nm for the films reported by other authors[9-12]. The functional dependence, of forming voltage and electric field, on thickness of switching layer and work function of top electrode is plotted in Fig. 3(d).There appears to be a stronger dependence on metal work function of top electrode than the thickness of the SiC film. This behaviour is similar to that observed in several oxide based RRAMs[1-6]. It is also observed that the forming field, which is the forming voltage divided by the thickness, decreases with thickness of SiC as predicted by Dearnaley etal[17].

In addition to the forming voltage,Set/Reset currents and voltages are also dependent on the switching layer thickness and the top electrode material, as is evident from Fig. 3(a)-(c) for devices D1-D7. Fig. 4(a)-(c)shows the endurance data for devices D2,D5 and D7 which have the same switching layer thickness but different work functions for the top layer.This data indicates that the Set and Reset voltages are increasing and the Reset current is decreasing with increase in work function of top electrode. The endurance of the devices D1,D3,D4,D6 is shown in Fig. S2(a)-(d). The range of set and Reset voltages for diffeent devices are summarized in Table 1.

| Device Name | Set voltage Range(V) | Reset voltage Range(V) |
|---|---|---|
| D1 | 0.5-1.27 | 0.12-1 |
| D2 | 0.5-1.5 | 0.1-1.5 |
| D3 | 0.34-2.25 | 0.14-1.9 |
| D4 | 0.3-2.73 | 0.17-4.5 |
| D5 | 0.4-3.15 | 1-3 |
| D6 | 0.3-3.84 | 0.35-4.4 |
| D7 | 0.8-5 | 0.34-5 |

**Table 1.**Range of Set and Reset voltages for devices D1-D7

Devices D1, D4 and D6, all of which have 50nm as the switching layer thickness, give consistent results for less than 6 cycles. For D5 and D7 with 100nm switching layer thickness and top electrode Cu and Pt respectively, give distinct ranges for Set and Reset resistances even over 40 cycles.

High $R_{OFF}/R_{ON}$ ratios are particularly necessary for logic devices to have a clear distinction between the ON and OFF states[18]. In the present study, the $R_{OFF}/R_{ON}$ ratio is in the range of $10^8$-$10^2$ for Cu based devices and $10^6$-$10^2$ for the Pt based devices with SiC thickness of 100 nm and $10^4$ and $10^3$ when the thickness is 50 nm. Morgan et al[12] and Zhong et al[10] have observed $R_{OFF}/R_{on}$ ratio of $10^7$ - $10^8$ in Cu/a-SiC/Au devices. In contrast, Lee et al[11] reported ratios of the order of $10^3$ in a Cu/SiC/Pt device. The values obtained in the present study are thus comparable to those reported in the literature. Significantly, the devices in the current study are unipolar in nature, in contrast to the earlier work where co-existence of unipolar and bipolar switching has been observed. Unipolar switching is considered a better option than bipolar switching for some applications since a single polarity pulse provides the possibility of unipolar diodes for selection in an array which in turn results in simpler circuits[19]. It has been found in other systems that unipolar switching can be attributed to localized Joule heating effects leading to the formation of conductive filaments at the formative voltage[7,8,19]. The switching is then due to the formation and rupturing of the filaments, depending on the voltage. However, in the case of Cu/SiC/Pt where non-polar switching is observed, it is postulated that Cu filaments form due to electrochemical reactions[3-7]. In the Set process there is anodic dissolution of Cu under positive bias. The $Cu^{2+}$ cations are driven by the field towards the bottom (Au) electrodes where they reduce to Cu atoms forming conductive filaments. In this model the filament rupture in the OFF state is due to electrochemical dissolution of the Cu filaments accompanied by local Joule heating[10,11].

Endurance of the devices D2, D5 and D7, with the best performance are shown in Fig.4(a), 4(b) and (c), respectively. High Resistance state (HRS)/Low Resistance State (LRS) values for these devices are in the order of $10^4$, $10^6$ and $10^7$ respectively. The endurance for another set of devices are shown in Fig.S2 (e) and (f), from which it is clear that, for the first 10 cycles the devices are quite stable and exhibit HRS/LRS ratios of $10^6$ and $10^4$ for D5 and D7 respectively. On further cycling there is a drastic decrease in the HRS/LRS ratio with the device eventually failing at around 50 cycles, as evidenced by the HRS/LRS ratio approaching 1. The retention time for the devices D2, D5 and D7 checked at 0.1V(500μA), is greater than $10^4$ sec as observed from Fig.5(a)-(c). There are fluctuations in the HRS and LRS values which can be attributed to the surface morphology being non-uniform.

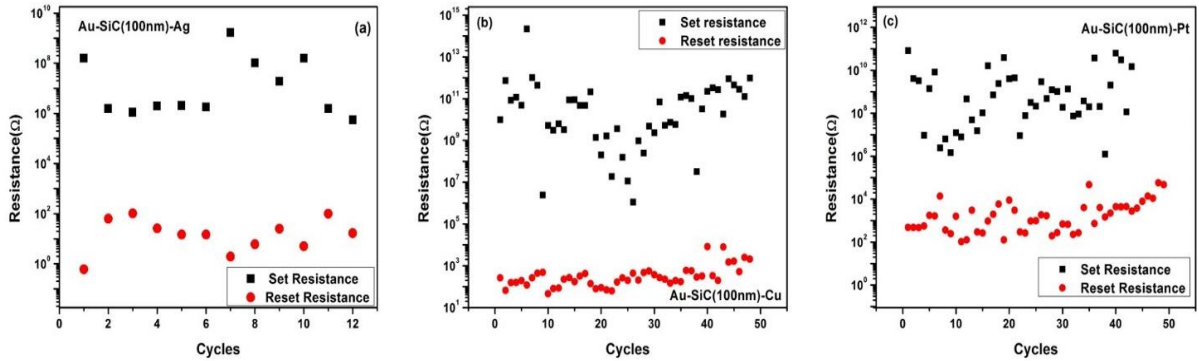

**Figure 4.** Endurance of (a) Au-SiC(100nm)-Ag, (b) Au-SiC(100nm)-Cu, (c) Au-SiC(100nm)-Pt based devices.

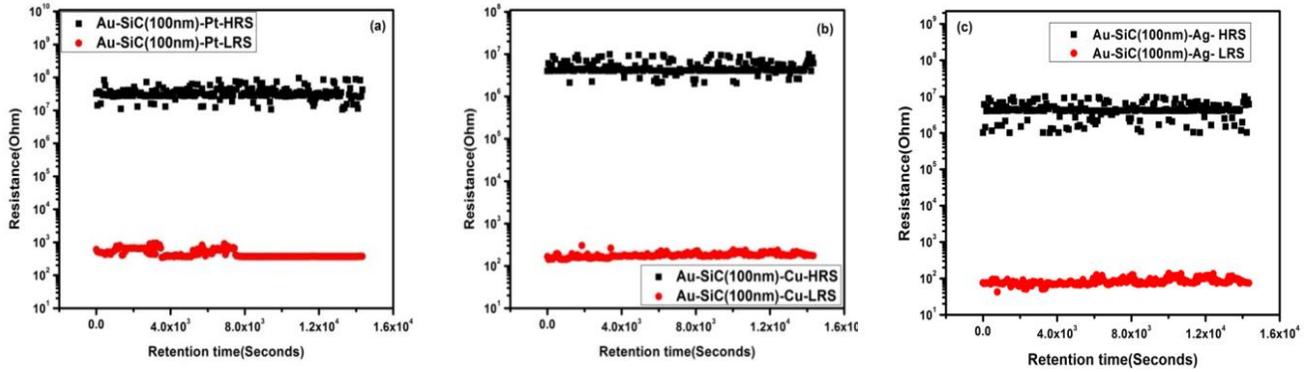

**Figure 5.** Retention time of (a) Au-SiC(100nm)-Pt, (b) Au-SiC(100nm)-Cu, (c) Au-SiC(100nm)-Ag based devices

The mechanisms of conductivity in the OFF and ON states are established by plotting double log I-V graphs as shown in Fig.6(a)-(f) and Fig. S3(a)-(h). For the devices D2, D5 and D7, with Ag, Cu and Pt as top electrodes respectively, conductivity in the HRS fits the model of thermionic emission across the metal/dielectric Schottky barrier in which current is varying from $10^{-12}$-$10^{-10}$A. In the current region of $10^{-9}$-$10^{-6}$A, space charge limited conduction(SCLC) in the low voltage region and Child's law ($I \sim V^2$) related conductivity in the high field region is observed shown in Fig. S4,S5 and S6. In the ON state, Ohmic conduction (high current region of $10^{-5}$-$10^{-3}$A) is observed. In contrast, for the devices D1, D4, D6 with the same electrodes and 50 nm thickness SiC, thermionic emission across the Schottky barrier appears to dominate in all the regions as shown in Fig. S3(a),(c),(e).

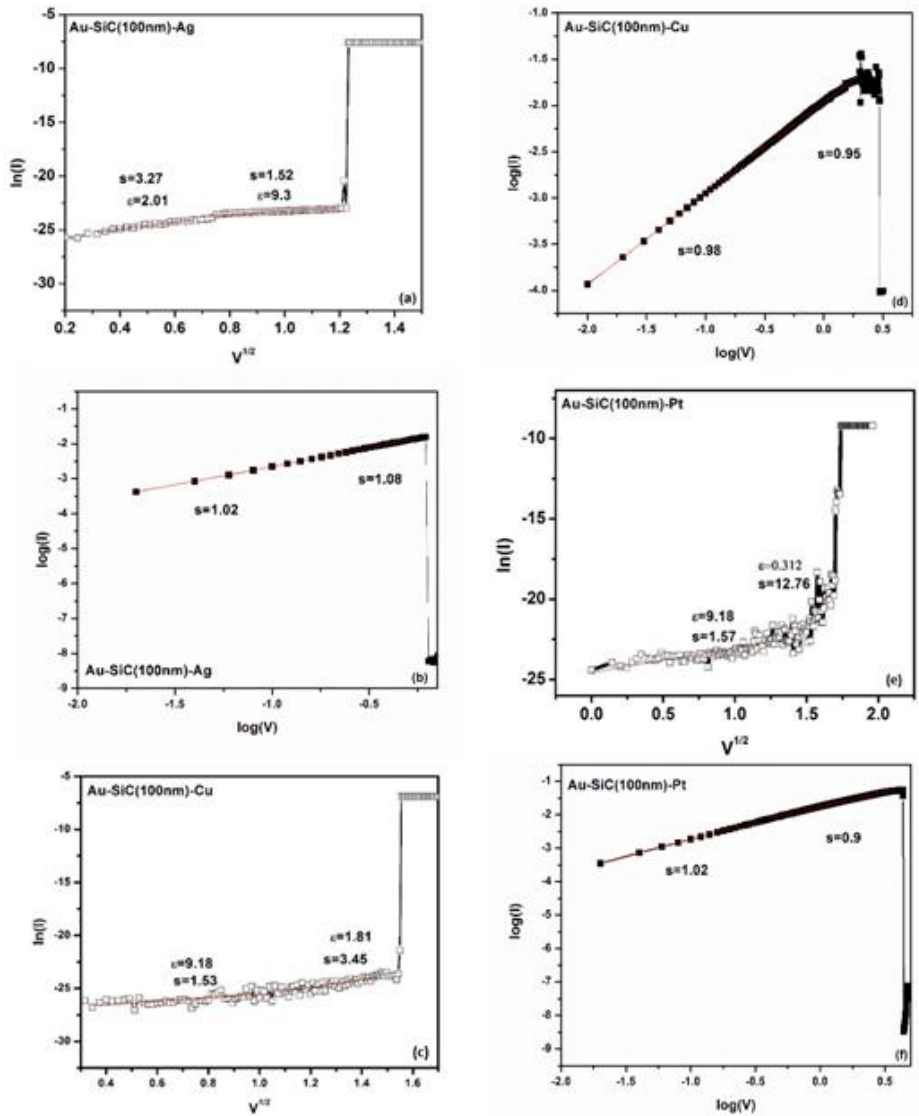

**Figure 6.** ln(I) vs V^0.5 and Double log I-V plots Curve fitted for Schottky emission and Ohmic conductivity as conduction mechanisms for the devices (a),(b) Au/SiC(100nm)/Ag, (c),(d) Au/SiC(100nm)/Cu, (e),(f) Au/SiC(100nm)/Pt in HRS and LRS respectively(where s is slope and ε is calculated dielectric constant from thermionic emission equation)

There seems to be consensus in literature that the origin of switching in SiC devices is related to the formation of Cu filaments[9-12]. However, the fact that the device Au/SiC/Pt with Pt electrode (wherein both bottom and top electrodes are both electrochemically inert) also exhibits switching, points to the possibility of other mechanisms contributing to resistive switching. Pt is electrochemically inert and, therefore, electrochemical dissolution is not possible. The high $R_{OFF}/R_{ON}$ ratios and the occurrence of Joule heating indicates the possibility of localized melting and dissociation of SiC.

To investigate this, the XRD patterns of the tested devices D5 and D7 were recorded and are shown in Fig. 7(a) and (b). Remarkably, all the peaks corresponding to Cu, Pt and Au as seen in fig 1(a) and (b) disappeared after testing. The diffraction pattern clearly indicates the interdiffusion of the different layers in the device structure. Cross-sectional SEM images of these devices before and after testing are shown in Fig. 8 (a) and (b) respectively. It is evident from these images that interfaces between the different layers can be clearly identified (as given by the lines) prior to testing. However, after a few cycles of testing, in the case of D5, there is inter-diffusion between the anode, cathode and active layer. In fact, in the case of D5 (Copper as Top Electrode), filaments in the shape of nanoneedles/ filaments protruding out of the top surface are observed. Another significant feature is that the columnar microstructure that was visible for the untested device is completely changed in the tested device. In the case of D5, the top and bottom electrodes transform into cylindrical nano-pancake like structures stacked vertically along the surface. Filaments, aligned in the direction of the applied field, appear on the top electrode. The inter-diffusion would indicate that melting of SiC accompanied by dissociation is taking place in the presence of the applied field. It is known that bulk SiC is stable upto temperatures >2000°C and only at temperatures greater than this, dissociation occur[20]. Since there is evidence for Joule heating it is possible that the process of switching is initiated by the formation of Cu nanofilaments in the first few cycles. Thereafter, as the Cu ions are transported through the dissociating layer of SiC they react with the Si to form a silicide. Similar is the case with Pt which is also known to readily form silicides[21-24]. Both these metals do not form carbides and it is, therefore, expected that the carbon would segregate to the surface.

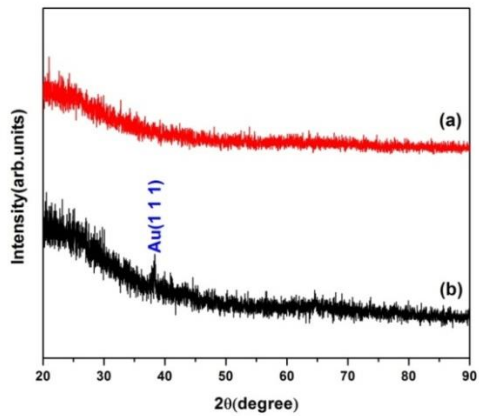

**Figure 7.** X-ray diffraction patterns of the devices (a) Au/SiC(100nm)/Pt, (b) Au/SiC(100nm)/Cu after the application of electric field for endurance test

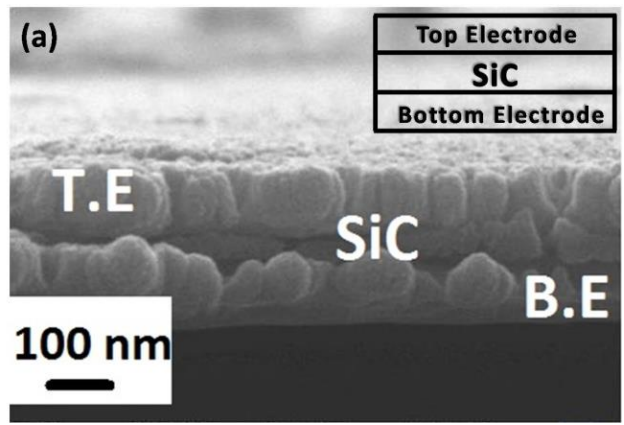

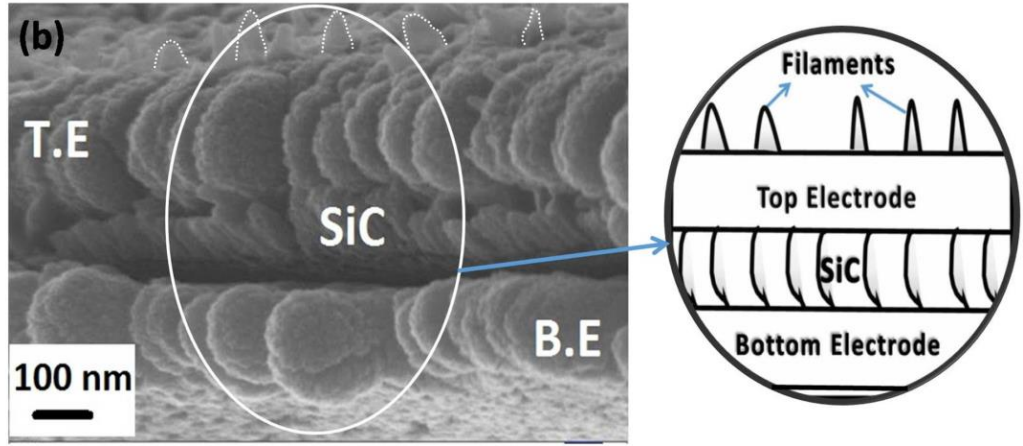

**Figure 8.** FESEM images of device (a) which is not cycled for RRAM testing with inset showing schematic of crosssectional view of device, (b) after cycling for few cycles with marked portion containing crossectional view of this device containing filaments

The Raman spectra of the Au/SiC/Pt and Au/SiC/Ag devices after a few cycles of switching are shown in Fig. 9(a) and (b). In the Raman spectra we can clearly observe the characteristic in-plane vibration modes (G), defect related peak (D) and second order vibration modes (2D) peaks of graphitic material. Table 2 below summarizes the peak positions and FWHM of the Raman peaks.

| Device Type | Raman peaks | Position (cm$^{-1}$) | FWHM (cm$^{-1}$) |
|---|---|---|---|
| Au/SiC/Pt | D | 1341 | 187 |
|  | G | 1596 | 77 |
|  | 2D | 2690 | 34 |
| Au/SiC/Ag | D | 1361 | 315 |
|  | G | 1581 | 96 |
|  | 2D | - | - |
| Graphite | D | 1350 |  |
| (from Ref[25]) | G | 1580 | 13 |
|  | 2D | 2700 | 25 |

**Table 2.** Peak positions and FWHM of Rman peaks of devices Au/SiC/Pt, Au/SiC/Ag in comparison with Graphite from Ref.25

The G mode and 2D mode peak position in few layer graphene stronglydepends on strain and chemical doping due to local environment (substrates, adsorbates). The position of G peak in Au/SiC/Pt is similar to that reported for graphene on 6H- SiC(0001)while for Au/SiC/Ag device it is similar to those reported for graphite[26,27]. The presence of large defect activated D peaks in all the decives suggest that a large fraction of the carbon might be $sp^3$ bonded carbon. The line shape of 2D peak depends on the number of layers present in few layer graphene. For a single layer of graphene on $SiO_2$ substrate the FWHM of 2D peak is 34 cm$^{-1}$. In addition there are other defect related modes like D', 2D' (overtone of D'), D+D'. In Au/SiC/Pt , the broad peak at high wavenumber has been deconvoluted into 2D, 2D', and D+D' components which gives 2D peak position at 2690 cm$^{-1}$ as shown in Figure S7. However, in Au/SiC/Ag this peak is so broadened that it cannot be reliably deconvoluted.

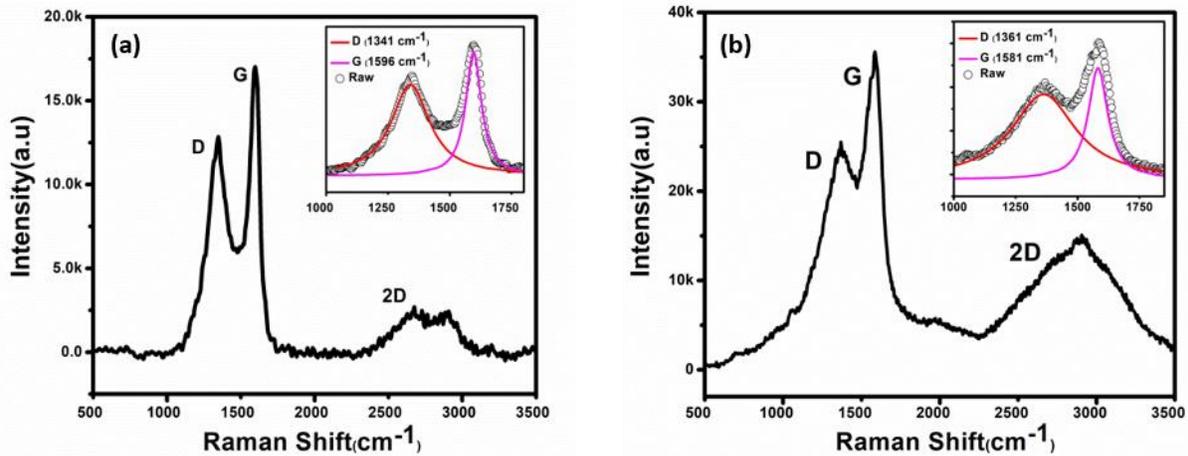

**Figure 9**. The Raman spectra of devices(a)Au-SiC-Pt (b)Au-SiC-Ag with the inset showing the D and G peak after fitting with Lorentzian functions

The Raman peak shape is highly influenced by disorder in the material. As the grain size decreases to nanocrystalline range , the disorder increases which leads to broadening of Raman peaks[28,29]. In all the devices, Raman peaks are broadened, indicating the presence of nanocrystalline graphite consistent with the FESEM images shown in Fig. 8.It is to be noted that graphite peaks are not observedin the case of untested devices. Further confirmation of the graphite presence is obtained through XPS measurements, shown in Fig.10. The high resolution XPS spectra of the C1s peak could be deconvoluted into 7 components at 283.3 eV (SiC), 284.6 eV (C=C), 285.1 eV (C-C), 285.8 eV (C-OH), 286.0 eV (C-O), 286. 9 eV (O-C-O) and 288.5 eV (C=O). Both $sp^2$ and $sp^3$ bonded carbon are present. However, most of the most of the carbon is bonded to either OH or other oxygen containing groups detected by XPS.

It is evident from these studies that, in the case of electrochemically inert top electrodes (Pt and Ag)unambiguous presence of nanocrystalline graphitic filaments is established. However, in the devices with Cu TE, the same mechanism cannot explain conduction, since a very small amount of graphite is present. It is known that Cu has the lowest redox potential among Ag and Pt, so it willreadily start to dissolve at much lower potential. As a consequence, it is more likely conduction is due to metallic Cu filaments in these devices.

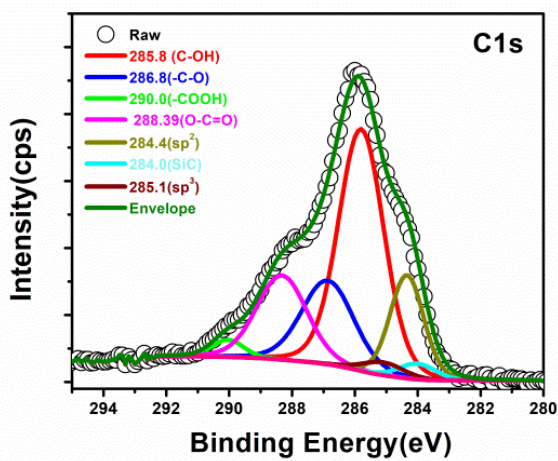

**Figure 10.** C1s core level high resolution XPS spectra.The C1s peak has been deconvoluted into seven components corresponding to different functional groups

The eventual failure of the devices can either be due to the formation of metallic conducting filaments or formation of nanocrystalline graphitic channels. One of the two competing mechanisms will dominate depending on the redox potential of the metal electrode.

**SUMMARY**

Resistive Random access memories based on amorphous SiC thin films are deposited using a single target Co-sputtering technique andwith electrodes Pt-Au,Ag-Au and Cu-Au are studied.The role of SiC film thickness and top electrode material (Cu, Pt and Ag) in determining switching performance, is examined. The existence of an optimal thickness of SiC (100 nm) for the best endurance is demonstrated. The most important observation of this work is to establish the formation of nanocrystalline graphite through Raman and x-ray photoelectron spectroscopy studiesfor Pt and Ag top electrodes. It is demonstrated thatwhen the top electrode is electrochemically inert (Pt,Ag) switching is accompanied by formation of graphitic filaments, while a different switching mechasim - likely to be formation of metalic filaments dominates when the top electrode(eg.Cu) is elecrochemically active.

**Acknowledgements**

The authors acknowledge facilities provided by the Central Facility for Nanotechnology, UGC-UPE and DST-PURSE programmes of University of Hyderabad. A portion of this research was performed using facilities at CeNSE, funded by Ministry of Electronics and Information Technology (MeitY), Govt. of India, and located at the Indian Institute of Science, Bengaluru.

# Resistive switching behaviour of amorphous silicon carbide thin films fabricated by a single composite target magnetron sputter deposition method

P.Chaitanya Akshara[1], Nilanjan Basu[2], Jayeeta Lahiri[2], G.Rajaram[1,2] and M.Ghanashyam Krishna[1,2,*]

[1]Centre for Advanced Studies in Electronics Science and Technology, [2]School of Physics, University of Hyderabad, Hyderabad-500046, Telangana, India
*Corresponding Author: email: mgksp@uohyd.ernet.in

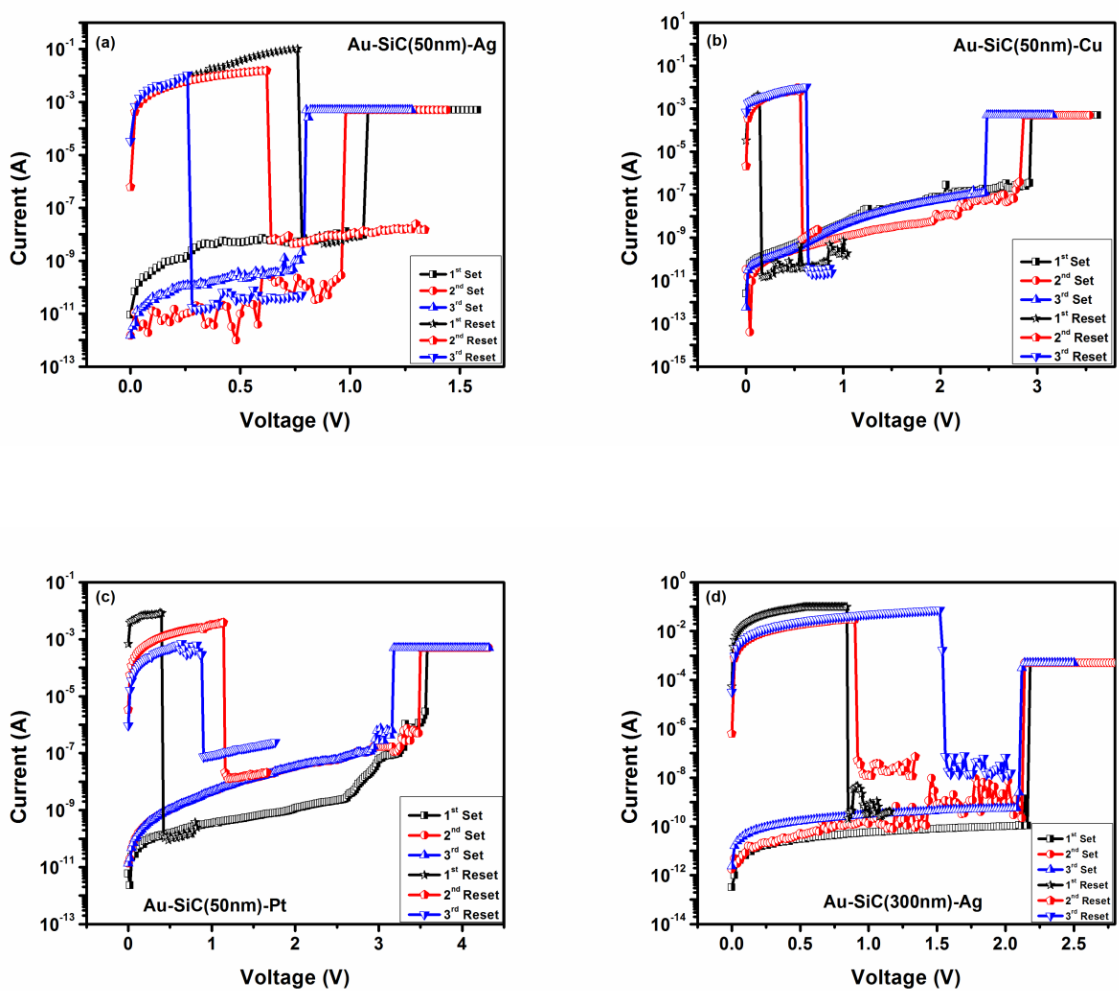

**Figure S1.** Switching characteristics of devices in the first 3 cycles (a) Au/SiC(50nm)/Ag, (b)Au/SiC(50nm)/Cu, (c) Au/SiC(50nm)/Pt,(d)Au/SiC(300nm)/Ag

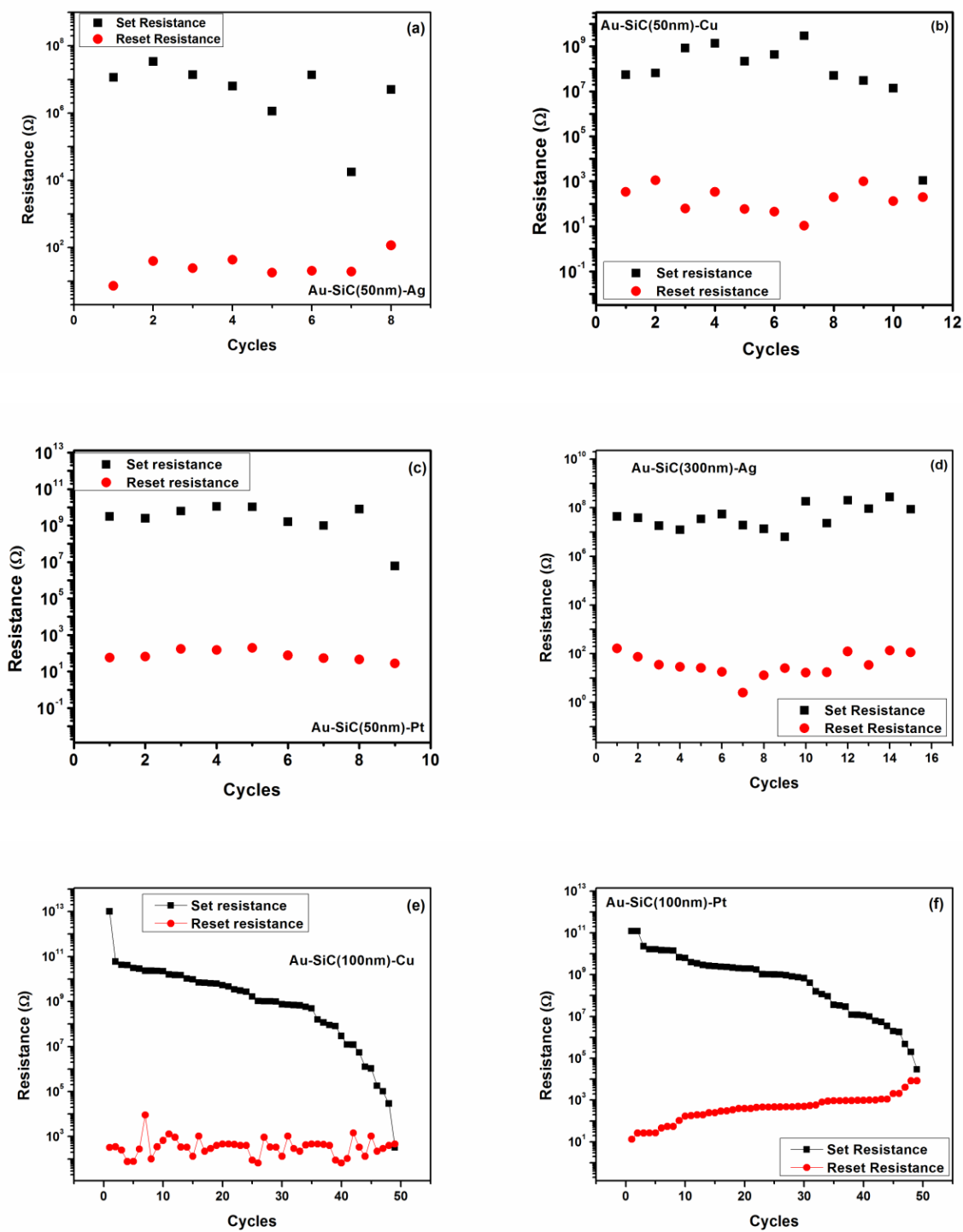

**Figure S2.** Endurance of devices (a) Au/SiC(50nm)/Ag, (b) Au/SiC(50nm)/Cu, (c) Au/SiC(50nm)/Pt, (d) Au/SiC(300nm)/Ag, (e) Au-SiC(100nm)-Cu, (f) Au-SiC(100nm)-Pt

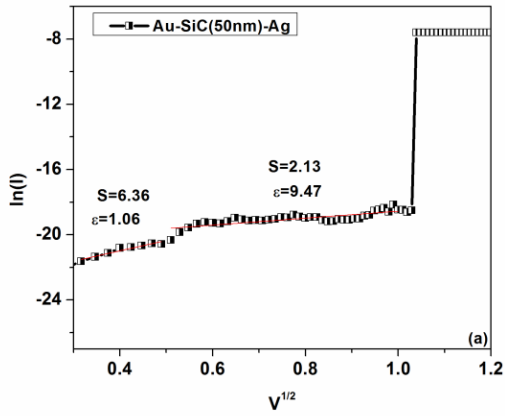
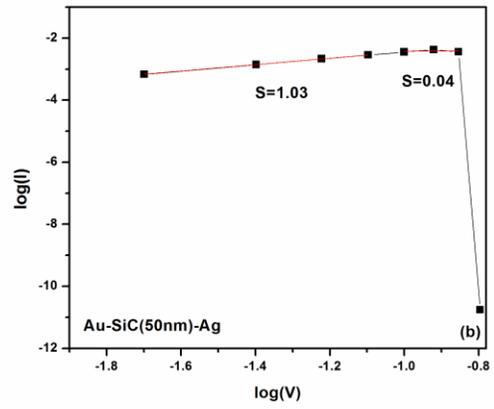
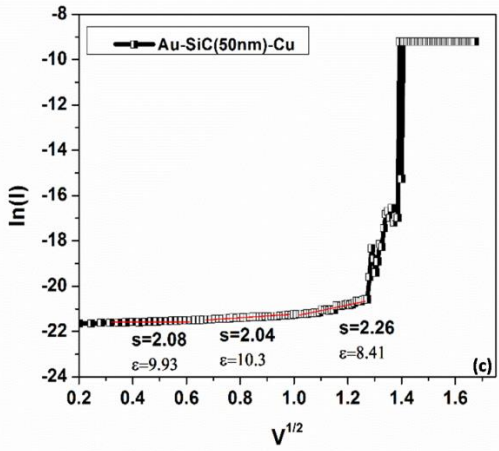
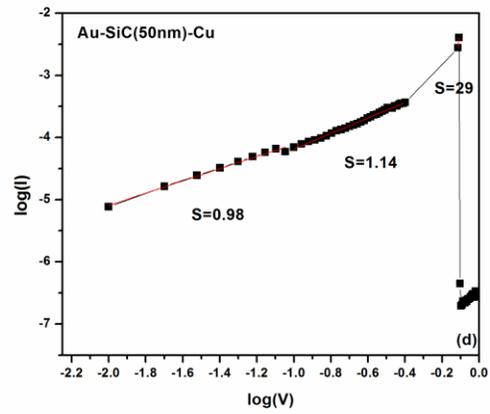
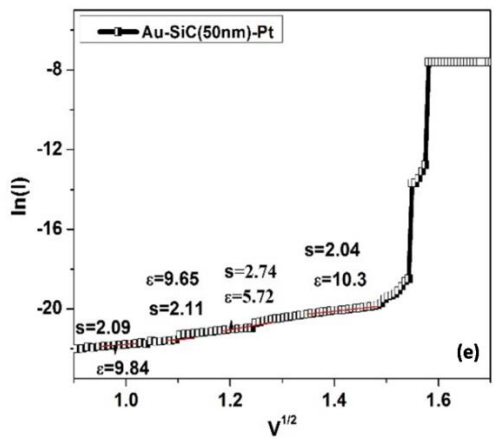
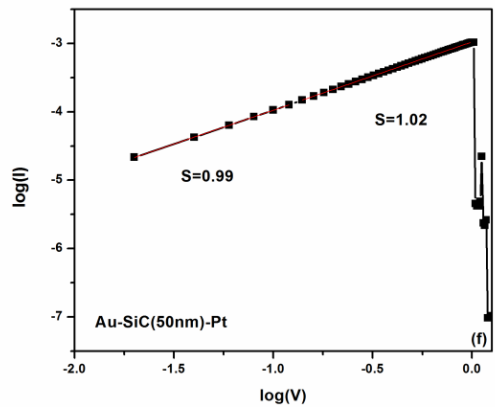

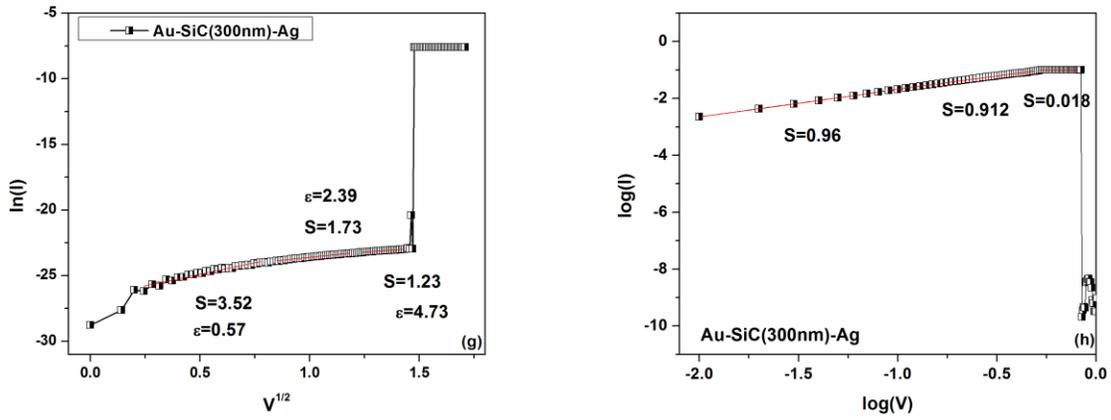

**Figure S3.** ln(I) vs V^0.5 and Double log I-V plots Curve fitted for Schottky emission and Ohmic conductivity as conduction mechanisms for the devices(a),(b) Au/SiC(50nm)/Ag, (c),(d) Au/SiC(50nm)/Cu, (e),(f) Au/SiC(50nm)/Pt,(g),(h) Au/SiC(300nm)/Ag in HRS and LRS respectively(where s is slope and ε is calculated dielectric constant from thermionic emission equation)

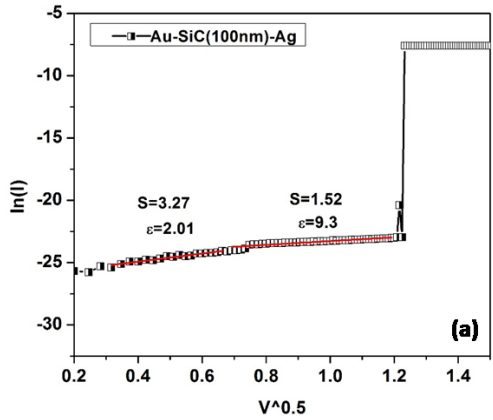
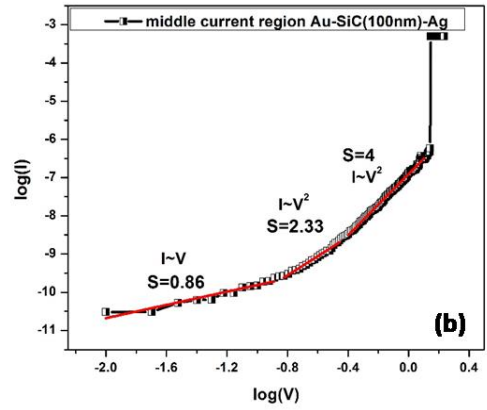
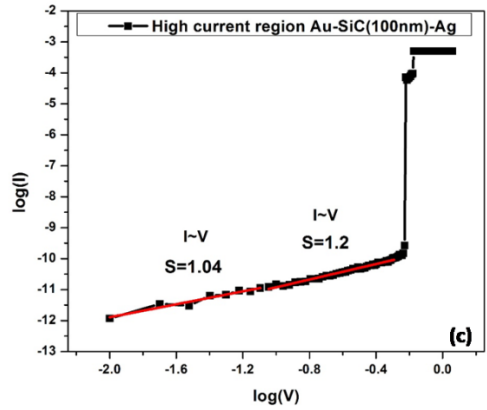

**Figure S4.** Conduction mechanism for the device Au-SiC(100nm)-Ag in different current regions curve fitted for ohmic, schottky mechanism equations

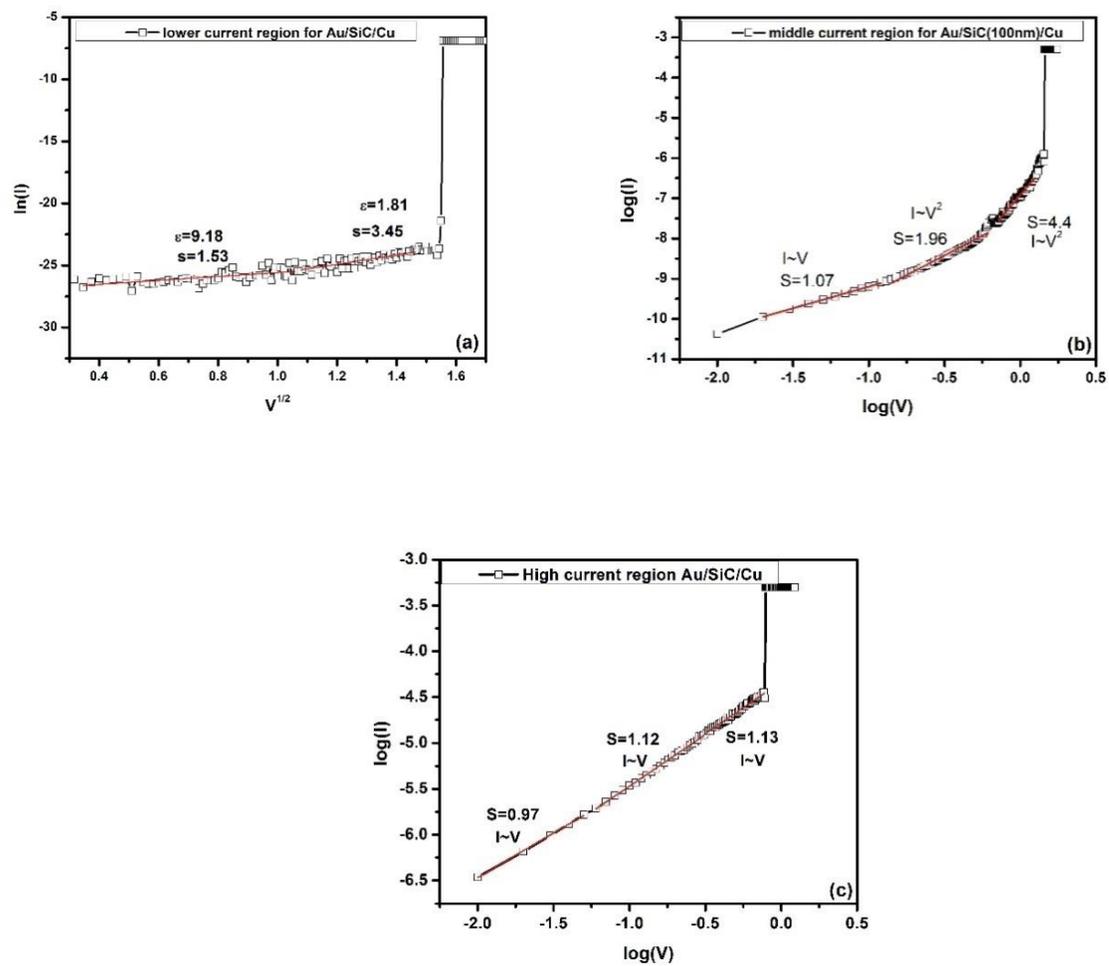

**Figure S5.** Conduction mechanism for the device Au-SiC(100nm)-Cu in different current regions curve fitted for ohmic, schottky mechanism equations

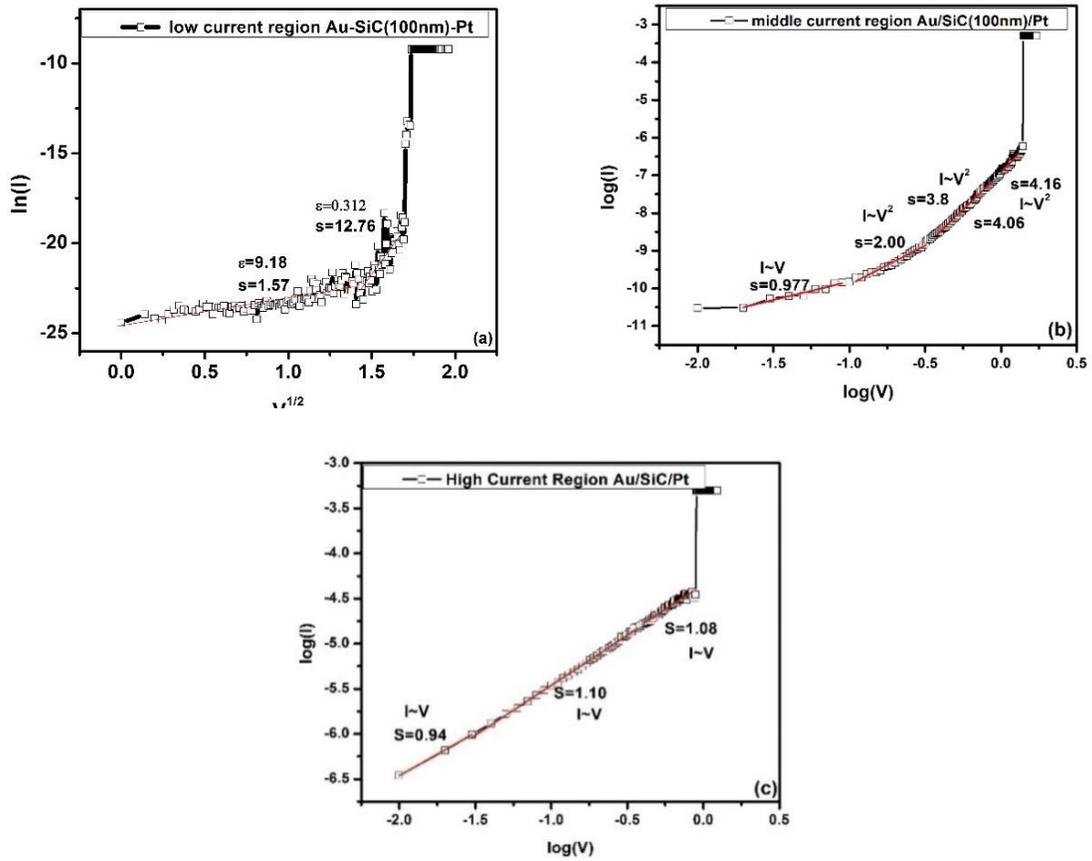

**Figure S6.** Conduction mechanism for the device Au-SiC(100nm)-Pt in different current regions curve fitted for ohmic, schottky mechanism equations

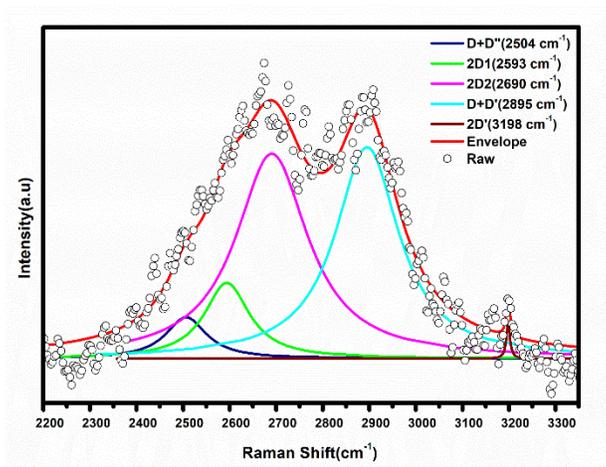

**Figure S7.** The Raman spectra of device Au-SiC-Pt peak position at 2690 cm$^{-1}$ (2D) deconvoluted into 2D, 2D', and D+D' components